\documentclass{emulateapj}
\usepackage{amssymb,amsmath}
\usepackage{graphicx}
\usepackage{natbib}
\begin{document}
\title{Binary YORP and Evolution of Binary Asteroids}
\author{Elad Steinberg\altaffilmark{1} and Re'em Sari\altaffilmark{1,2}}
\altaffiltext{1}{Racah Institute of Physics, Hebrew University, Jerusalem 91904, Israel}
\altaffiltext{2}{Theoretical Astrophysics, Caltech 350-17, Pasadena, CA 91125, USA}

\email{elad.steinberg@mail.huji.ac.il}
\begin{abstract}
The rotation states of kilometer sized near earth asteroids are known to be affected by the YORP effect.
In a related effect, Binary YORP (BYORP) the orbital properties of a binary asteroid evolves under a radiation effect mostly acting on a tidally locked secondary.
The BYORP effect can alter the orbital elements in $\sim 10^{4-5}$ years for a $D_{p}=2\;km$ primary with a $D_{s}=0.4\; km$ secondary at $1\; AU$.
It can either separate the binary components or cause them to collide. In this paper we devise a simple approach to calculate the YORP effect on asteroids and BYORP effect on binaries including $J_2$ effects due to primary oblateness and the sun.
We apply this to asteroids with known shapes as well as a set of randomly generated bodies with various degrees of smoothness.
We find a strong correlation between the strengths of an asteroid's YORP and BYORP effects. Therefore, a statistical knowledge on one, could be used to estimate the effect of the other.
We show that the action of BYORP preferentially shrinks rather than expands the binary orbit and that YORP preferentially slows down asteroids. This conclusion holds for the two extremes of thermal conductivities studied in this work and assuming the asteroid reaches a stable point, but may break down for moderate thermal conductivity.
The YORP and BYORP effects are shown to be smaller than what could be naively expected due to near cancellation of the effects on small scales.
Taking this near cancellation into account, a simple order of magnitude estimate of the YORP and BYORP effects as function of the sizes and smoothness of the bodies is calculated. Finally, we provide a simple proof showing that there is no secular effect due to absorption of radiation in BYORP.
\end{abstract}

\section{Introduction}
When NEA orbit around the sun they are constantly subjected to the sun's radiation.
In equilibrium, the total energy absorbed by the asteroid must be re-emitted. Yet, asymmetry in the asteroid's geometry results in
a residual force that tends to change the asteroid's motion \citep{Rub2000}.
This residual force, called the YORP effect, can significantly change the spin rate of a kilometer sized asteroids at a distance of $1\; AU$ from the sun in $10^5-10^6$ years.
The YORP effect has been successfully measured for several NEA \citep{YORPchange, apollochange, geochange}. The high abundance of fast rotators in the NEA family \citep{pravec2008} might be explained by the YORP effect. A rubble pile NEA might undergo fission if the YORP effect accelerates it beyond it's rotational breakup velocity \citep{VC02}; this mechanism might explain the formation of binary NEA.

The differential acceleration between the two components of the binary is mostly due to the acceleration of the secondary due to its larger surface to volume ratio.
A coherent effect also requires at least one of the components to be tidally locked which is more common for the secondary.
The net radiation force acting on the secondary will produce a torque relative to the primary. This Binary YORP (BYORP) effect, first suggested by \cite{CB05}, evolves the orbit of the binary on fairly fast timescales ($\sim 10^5$ years for a $D_{p}=2\;km$ primary with a $D_{s}=0.4\; km$ at $1\; AU$ separated by $a=1.5D_{p}$).
The spin rate and the obliquity of the asteroid evolve due to the YORP effect.
Similarly, the semi-major axis of the binary and its inclination relative to the orbital plane around the sun evolve by the BYORP effect.
In addition the BYORP effect changes the eccentricity vector of the binary.
We show that there are preferred end states for both effects. YORP tends to slow the spin rates, while BYORP tends to shrink the semi-major axis in the binary case.

In \S2 we introduce our model method and assumptions.
In \S3 we show that neighboring areas on the surface of the asteroid tend to have opposing effects.
We provide order of magnitude estimates of the YORP effect as function of the sizes of the body and
its smoothness. A method of precise calculation, as well as counting for thermal lag due to finite
thermal inertia is shown in \S4 and the results are discussed in \S5.
Analogous results for the BYORP effect are derived in \S6.

\section{Structure Models and Coordinate Systems}
\subsection{Modeling Method}
We model the asteroids by means of tessellation, where the asteroid's surface is described by a set of triangles.
We neglect the effect of shadows cast by one facet over another and we assume that the emission from each point on the surface of the asteroid is isotropic.
Since the orbital period of the asteroid around the sun is much longer than its rotation time, it is assumed that there are no resonances between the asteroid's orbit around the sun and the asteroid's revolution around itself.
Constant density is assumed, although possible effects of density non uniformity will be briefly addressed.

\subsection{Coordinate Systems}
\label{sec:cor}
Two coordinate systems will be used throughout this paper:
\begin{enumerate}
    \item An inertial frame with axes labeled $x,y,z$. The $z$ axis is perpendicular to the orbital plane of the asteroid around the sun.
    The x axis is chosen to coincide with the projection of the asteroid's spin vector on the $xy$ plane.
    The origin is chosen to be the sun.
    This system will be referred to as the inertial system.
	\item The principle axes of the asteroid, labeled $x',y',z'$, where $z'$ is parallel to the spin vector. This system will be referred to as the asteroid system.
\end{enumerate}

\section{Scaling of the Radiative Torque}
\label{sec:em}
In this section we discuss the net acceleration and total torque that arises from the emission of radiation by
a body of arbitrary shape. From symmetry, spherical bodies do not exhibit torque or acceleration, and we derive here the general scaling of the torque and acceleration as function of the roughness of the body, or its deviations from sphericity.
 We treat only the effect that arises from emission of radiation since \cite{absor} have shown that there is zero secular change due to absorption of radiation.

We assume that the orbital time around the sun and its rotational period are non commensurate, and we can therefore perform the time average by averaging over the orbit and the spin sequentially. We find it more convenient to first fix the angle of rotation of the asteroid around its axis and average the torque applied to the asteroid during an orbit around the sun, and then average the torque as the asteroid revolves around its axis.

In order to calculate the effect of YORP, the complete structure of the asteroid needs to be known. In this section we consider the scaling relations of YORP and provide order of magnitude estimate of the effect.

The change in the spin rate, $s$, of a homogeneous asteroid scales according to: \begin{equation}
    \dot{s}\sim\frac{\Phi}{\rho R^2}
\end{equation}
where $\Phi$ is the solar radiation momentum flux given by
\begin{equation}
\Phi\equiv \frac{L_\odot}{4\pi cd^2\sqrt{1-e^2}}.
\end{equation}
 Here $L_{\odot}$ is the solar luminosity, $e$ is the eccentricity, $c$ is the speed of light, $d$ is  the orbital semi-major axis, $\rho$ is the density, and $R$ is the length scale of the asteroid. The eccentricity dependence arises from averaging the torque over a heliocentric orbit.

We construct simple models
to account for the asymmetry of the asteroid. We choose $n$ points randomly distributed on the unit sphere and connect them to create a tessellation of small triangles that encloses the asteroid (based on the Quickhull algorithm \cite{quickhull}, which produces $2n-4$ triangular facets for a given $n$).
This method of construction eliminates shadowing of one facet over another. For this body, we now calculate the radiation effects, and their scaling with $n$ or with the deviation of the shape of the body from a sphere. In the estimates below, we assume $n\gg 1$.

We define the deviation of the asteroid from a sphere with the same volume by comparing the normalized difference in their surface areas. \begin{equation}
    d_{s}\equiv\frac{S-4\pi r^{2}}{4\pi r^{2}}
\end{equation}
where $4\pi r^{3}/3$ is the volume of the asteroid and $S$ is the surface area of the asteroid. This definition of spherical deviation will be shown to be $d_{s}\propto n^{-1}$. In order to simplify the derivation we will assume that all of the facets are equilateral and that the center of mass (CM, hereafter) is located at the origin.
If we connect each one of the facets to the CM and thereby create a tetrahedron, the relation between $\theta$, which is the vertex angle of a tetrahedron face, and $\Omega\approx 2\pi/n$, the solid angle that the tetrahedron subtends, can be found by making use of L'Huilier's theorem:\begin{equation}
\theta^2=\frac{4\Omega}{\sqrt{3}}.
\end{equation}
The area of each facet is:\begin{equation}
S_j=\frac{\sqrt{3}}{4}4R^2\sin^2(\frac{\theta}{2})\approx\frac{\sqrt{3}}{4}R^2(\theta^2-\frac{\theta^4}{12})
\end{equation}
where $R$ is the radius of the sphere that covers the asteroid. The height of the tetrahedron is:\begin{equation}
h\approx R(1-\frac{\theta^2}{6})\label{eq:h}.
\end{equation}
The volume that is enclosed between the facet and the sphere is:
\begin{equation}
V\approx \frac{R^3\Omega-S\cdot h}{3}\approx \frac{\sqrt{3}R^3\theta^4}{48}=\frac{R^3\Omega^2}{3\sqrt{3}}.
\end{equation}
The total difference in the volume between the asteroid and the sphere is roughly $2n\cdot V$:
\begin{equation}
\frac{4\pi}{3}R^{3}-\frac{4\pi}{3}r^{3}\approx\frac{8\pi^2R^3}{3\sqrt{3}n}.
\end{equation}
The ratio between the radii is: \begin{equation}
\frac{r}{R}\approx 1-\frac{2\pi}{3\sqrt{3}n}.
\label{eq:rR}
\end{equation}
By equating the volume of the asteroid with the volume of a sphere with radius $r$ we have:
\begin{equation}
\frac{h}{3}\sum_{j=1}^{n}S_{j}=\frac{4\pi r^{3}}{3}
\end{equation}
and eq.\eqref{eq:rR} and eq.\eqref{eq:h} yield:
\begin{equation}
S=\sum_{j=1}^{n}S_{j}\sim 4\pi r^{2}(1+\frac{2\pi}{3\sqrt{3}n}).
\end{equation}
Therefore, we obtain: \begin{equation}
    d_{s}\propto\frac{1}{n} \label{eq:div}.
\end{equation}
The angle, $\gamma$, between the normal of a non-equilateral facet and the vector joining the sphere's center to the facet's centroid,
is of order $\theta$, so: \begin{equation}
\gamma\propto\frac{1}{\sqrt{n}}.
\end{equation}
The torque that is produced by a single facet is: \begin{equation}
\boldsymbol{\tau_j}\propto S_j\gamma_j\propto n^{-\frac{3}{2}}.
\end{equation}
Naively, we might expect that the total torque is a random walk summation over the individual facets, and therefore
\begin{equation}
\boldsymbol{\tau}_{naive} \cong \sqrt{n} \boldsymbol{\tau_j}   \propto\frac{1}{n}.
\end{equation}
However, the torques from neighboring facets are not independent. This dependence leads to a partial cancellation of the torque produced by neighboring facets that share a point.
The torque is given by
\begin{equation}
\boldsymbol{\tau}\propto(\mathbf{a}+\mathbf{b}+\mathbf{c})\times(\mathbf{b}\times\mathbf{a}+\mathbf{c}\times\mathbf{b}+\mathbf{a}\times\mathbf{c})\cdot\cos\lambda
\end{equation}
where $\lambda$ is the angle between the normal to the facet and the sun's flux, $\mathbf{a},\mathbf{b}$ and $\mathbf{c}$ are vectors from the origin to the vertexes located on the surface of the sphere. Consider a displacement of a single vertex, either radially or tangentially on the surface of the sphere, by an amount that would create a change of order unity in the torque, then the difference between the initial torque and the torque after displacing $\mathbf{a}$ by an amount $\mathbf{da}$ is:\begin{equation}
\begin{split}&\mathbf{da}\times(\mathbf{b}\times\mathbf{a}+\mathbf{c}\times\mathbf{b}+\mathbf{a}\times\mathbf{c})\cdot\cos\lambda\\
&+(\mathbf{a}+\mathbf{b}+\mathbf{c})\times((\mathbf{b}-\mathbf{c})\times\mathbf{da})\cdot\cos\lambda\\
&+(\mathbf{a}+\mathbf{b}+\mathbf{c})\times(\mathbf{b}\times\mathbf{a}+\mathbf{c}\times\mathbf{b}+\mathbf{a}\times\mathbf{c})\cdot d(\cos\lambda).
\end{split}\label{eq:toruqe_change}
\end{equation}
If we displace $\mathbf{a}$ by an angle $|\mathbf{da}|\sim n^{-\frac{1}{2}}$, then the first two terms in eq.\eqref{eq:toruqe_change} scale like $\mathcal{O}({n^{-\frac{3}{2}}})$ while the third term scales like $\mathcal{O}(n^{-2})$. Likewise, if we displace $\mathbf{a}$ radially by $|\mathbf{da}|\sim n^{-1}$ the second term will scale like $\mathcal{O}({n^{-\frac{3}{2}}})$ while the first and third term will scale like $\mathcal{O}(n^{-2})$.

Both the radial and tangential displacement of the vertex $\mathbf{a}$ caused a change in the torque of order unity ($\mathcal{O}({n^{-\frac{3}{2}}})$). However, the same vertex $\mathbf{a}$ is shared by several neighboring tetrahedrons. Summing the changes in the torque over all of the facets which share the vertex $\mathbf{a}$, we find that all contributions up to $\mathcal{O}({n^{-\frac{3}{2}}})$ cancel out and we are left with a contribution of $\mathcal{O}(n^{-2})$. The effective contribution of each such group will therefore scale like $\mathcal{O}(n^{-2})$.
A random walk process will now give:
\begin{equation}
    \boldsymbol\tau\propto d_{s}^{\frac{3}{2}}.
\end{equation}
Following \cite[GS hereafter]{RS2009}, we define $f_{Y}$ to be the ratio between the actual torque and the torque that would be exerted if all of the received radiation were emitted tangentially from the body's equator:
\begin{equation}
\boldsymbol\tau=\frac{2}{3}\pi R^3\Phi f_Y
\end{equation}
where the $2/3$ arises from assuming isotropic emission.
We therefore find:
\begin{equation}
    f_{Y}\sim d_{s}^{\frac{3}{2}}\label{eq:fy}.
\end{equation}

We have shown that the total YORP effect is comparable to the effect of a single facet. Therefore, the uncertainties in the asteroid's shape will induce errors in the torque estimate by an amount of:
\begin{equation}
\frac{\Delta \boldsymbol\tau}{\boldsymbol\tau}\approx\frac{\Delta R}{(R/N)}\approx\frac{\Delta R}{ d_s R}
\end{equation}
where $\Delta R$ is the physical length scale of the modeling error and $R$ is the length scale of the asteroid.

\section{Detailed Calculation \& Thermal Lag}
In order to compute the total average torque, we first compute the average torque arising from each facet's emission independently.
Let $\mathbf{n}$ be the unit vector pointing from the facet to the sun,
$\mathbf{ds}$ the vector perpendicular to the facet with a magnitude equal to its area and $\mathbf{r}$ the vector pointing from the
asteroid's CM to the centroid of the facet, then the torque of each facet due to Lambertian reflection while neglecting specular reflection is:
\begin{equation}\label{eq:torque1}
    \boldsymbol{\tau}_{reflect}=-\frac{2}{3}\frac{A\Phi d^2}{D^2}(\mathbf{n}\cdot\hat{\mathbf{ds}})(\mathbf{r}\times\mathbf{ds})
\end{equation}
where $D$ is the distance between the sun and the asteroid, $A$ is the asteroid's albedo that is assumed to be constant.
Thermal lag due to a finite thermal conductivity,$\kappa$, might influence the obliquity change rate (it can also reverse sign in extreme cases) but it hardly effects the spin change rate \citep{CV04}.
The effect of thermal lag can be estimated by having some constant temperature per facet, $T_{eq}$, and a time varying component, $\Delta T$, which lags in time relative to the solar insolation. A facet with a constant temperature produces a torque which is a constant in the asteroid frame. Averaging of this torque over the asteroid's revolution around itself leaves us only with the projection of the torque on the spin axis. Therefore, the contribution of the constant temperature to the obliquity term vanishes. The spin component has no preference to the phase at which the emission takes place, hence the spin component is unaffected by thermal conductivity.

As the thermal conductivity increases, the ratio $T_{eq}/\Delta T$ increases \citep{thermalLag}. For a typical asteroid, if $ \kappa \cdot s\gtrsim6\cdot10^{-5}\: W/(m\,s\,K)$ then the equilibrium temperature dominates over the time varying component. Recent studies have shown that some asteroids could have high enough thermal conductivity \citep{conductivity}. For these asteroids, one can neglect the obliquity term due to thermal re-emission. In this paper we explore two extremes which bound the possible behaviors, the first is the Rubincam approximation ($\kappa=0$) and the second is the high thermal conductivity regime.

The torque due to thermal re-emission is:
\begin{equation}\label{eq:torque2}
    \boldsymbol{\tau}_{emission}=-\frac{2}{3}\frac{(1-A)\Phi d^2}{D^{2}}({\mathbf{n}}\cdot{\hat{\mathbf{ds}}})(\mathbf{r}\times\mathbf{ds}).
\end{equation}
For the high thermal conductivity regime we evaluate only the spin component of this torque.

The general case of non-zero obliquity requires averaging the torque over both: \renewcommand{\labelenumi}{(\roman{enumi})}\begin{enumerate}
    \item The orbit of the asteroid around the sun.
	\item A revolution of the asteroid around itself.
\end{enumerate}

For a zero obliquity orbit, it is sufficient to average over either one of the above. Since we assumed that the spin period and the orbital period are non-commensurate, we can calculate these averages in arbitrary order. We find that it is more convenient to first average over the orbit around the sun, while holding each facet pointing in a fixed direction.
\subsection{Heliocentric Orbit Average}
The insolation, averaged over the orbit of the asteroid around the sun, is given by \citep{Ward74}\begin{equation}
<I>=\frac{1}{T}\int_{0}^{T}\frac{L_{\odot}}{4\pi cD^{2}}\mathbf{n}\cdot\mathbf{\hat{ds}}\,{d}t=\frac{\Phi\sin(\theta)}{\pi}
\end{equation}
where $T$ is the heliocentric orbital period and $\theta$ is the angle between $\mathbf{\hat{ds}}$ and the normal to the orbital plane.

\subsection{Asteroid Spin Averaging}
\label{euler}
Since we are interested in calculating the torque in the inertial system (system 1), we need to transform the torque from the asteroid system (system 3) to the inertial system. In the asteroid system the torque per facet is:
\begin{equation}
\begin{split}\label{eq:ds}
&\boldsymbol\tau'_j=\boldsymbol\tau'_{reflection,j}+\boldsymbol\tau'_{emission,j}=-\mathbf{r'}\times \frac{2S_j}{3}<I>\mathbf{ds'}\\
&=-\frac{2S_j}{3}<I>
	\begin{pmatrix}
    y'_j\cos\theta'_j-z'_j\sin\theta'_j\sin\phi'_j\\
	z'_j\sin\theta'_j\cos\phi'_j-x'_j\cos\theta'_j\\
	\sin\theta'_j(x'_j\sin\phi'_j-y'_j\cos\phi'_j)
\end{pmatrix}
\end{split}
\end{equation}
where $S_j$ is the facet's area, and $\theta'_j,\phi'_j$ are the polar and azimuthal angles accordingly.
The torque is then transformed into system 1 with the standard Euler angle rotation matrix: \begin{equation}
 \mathbf{R}(\psi)=\begin{pmatrix}\label{eq:rot}
-\cos\epsilon\sin\psi & -\cos\epsilon\cos\psi & \sin\epsilon\\
\cos\psi & -\sin\psi & 0\\
\sin\epsilon\sin\psi & \sin\epsilon\cos\psi & \cos\epsilon
\end{pmatrix}
\end{equation}
where $\psi$ is the rotation angle of the asteroid around itself and $\epsilon$ is the obliquity. For $\psi=0$ the asteroid system's $x'$ axis coincides with the $y$ axis of the inertial system.

The insolation in system 1 is:
\begin{equation}
\begin{split}
&\frac{\Phi\sin(\theta_j)}{\pi}=\frac{\Phi\sin(\cos^{-1}([\mathbf{R}\cdot\mathbf{\hat{ds}}]\cdot\mathbf{\hat{z}}))}{\pi}\\
&=\frac{\Phi[(\cos(\psi+\phi'_j)\sin\theta'_j)^2}{\pi}\\
&\frac{+(\cos\theta'_j\sin\epsilon-\cos\epsilon\sin(\psi+\phi_j)\sin\theta'_j)^2]^\frac{1}{2}}{\pi}.
\end{split}
\end{equation}

The averaged torque in the inertial system can now be rewritten as:
\begin{equation}\label{eq:torque3}
\boldsymbol\tau_j=-\frac{S_j\Phi}{3\pi^{2}}\int_{0}^{2\pi}\sin\theta_j(\mathbf{R}\cdot\boldsymbol\tau'_j)d\psi.
\end{equation}

\begin{figure}[t]
\epsscale{2}
\plottwo{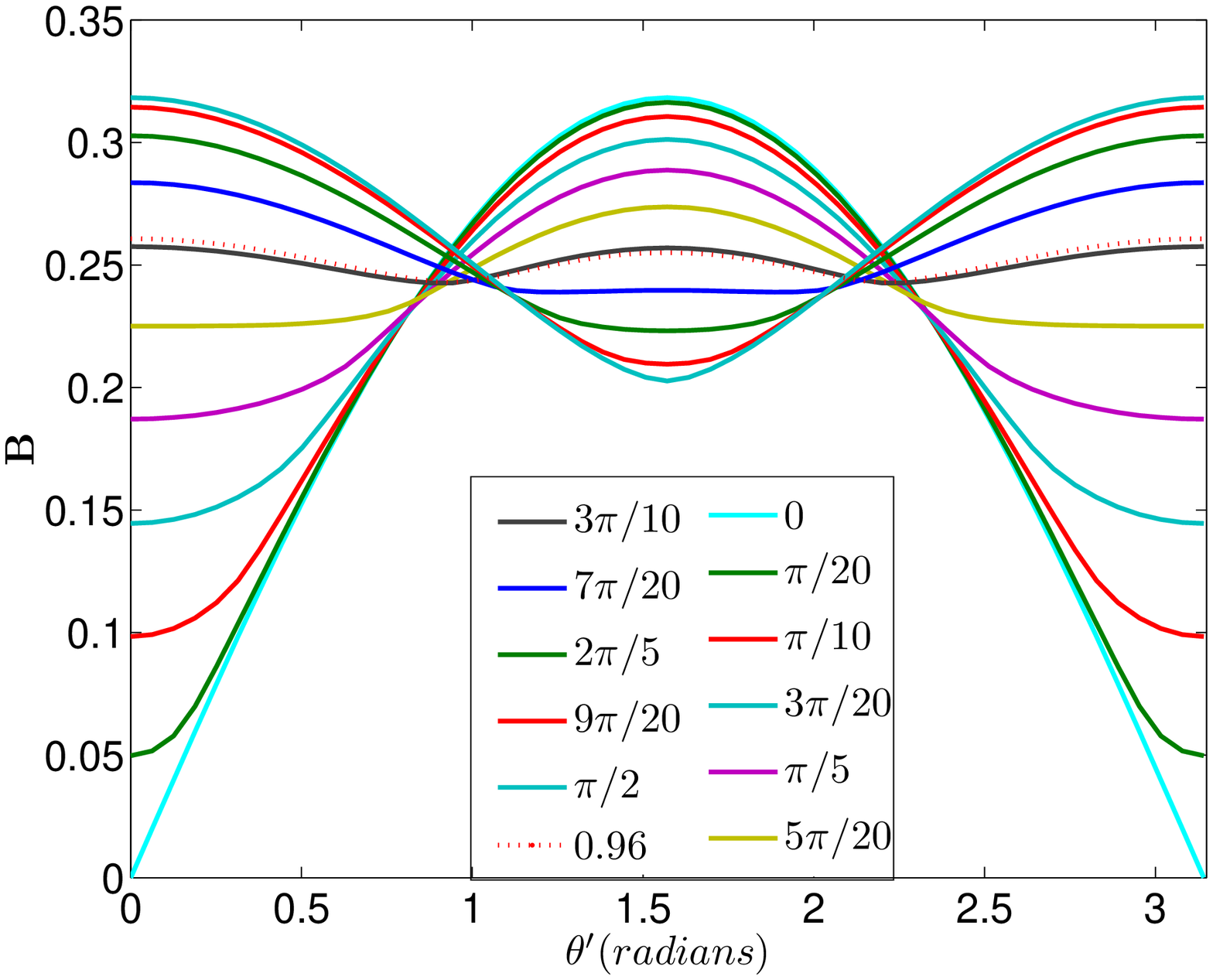}{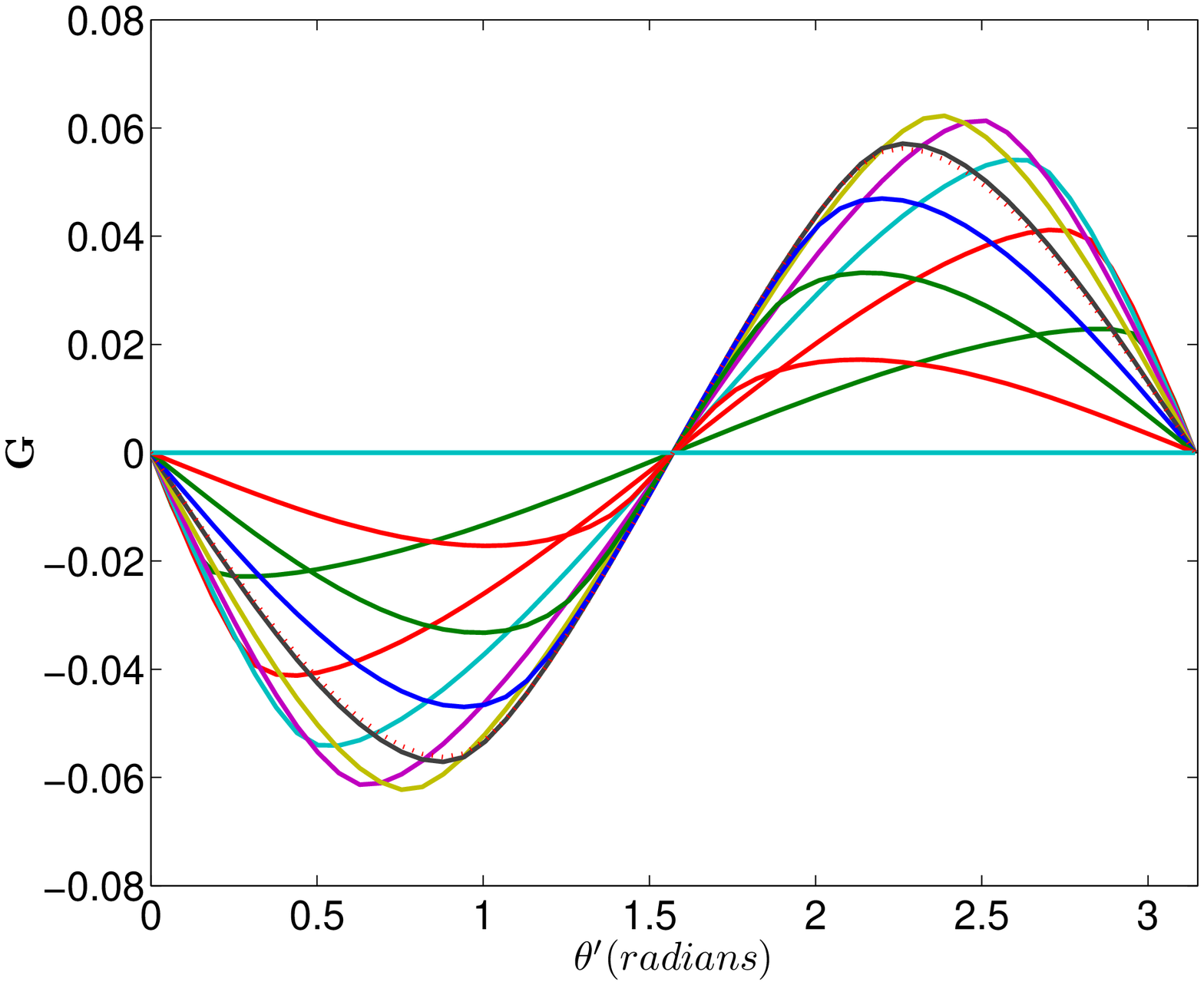}
\caption{The functions B and G for different obliquities as a function of the facet's polar angle relative to the spin axis. The dotted red line represents $\epsilon\approx 55^{\circ}$.}
\label{fig:numerical_figures}
\end{figure}

In order to calculate the torque we define the following numerical functions:
 \begin{equation}
 \begin{split}
 &B(\theta',\epsilon)\equiv\frac{1}{2\pi^{2}}\int_{0}^{2\pi}[(\cos\psi\sin\theta')^{2}\\
 &+(\cos\theta'\sin\epsilon-\cos\epsilon\sin\psi\sin\theta')^{2}]^\frac{1}{2}{d}\psi
 \label{eq:I}
 \end{split}
 \end{equation}
 and,
 \begin{equation}
 \begin{split}
&G(\theta',\epsilon)\equiv\frac{1}{2\pi^{2}}\int_{0}^{2\pi}\sin\psi[(\cos\psi\sin\theta')^{2}\\
&+(\cos\theta'\sin\epsilon-\cos\epsilon\sin\psi\sin\theta')^{2}]^\frac{1}{2}{d}\psi \label{eq:G}.
\end{split}
\end{equation}
Notice that the physical interpretation of the function $B$ is simply the average insolation a facet receives at a given angle $\theta'$. This result is similar to the one derived in \cite{Ward74} and \cite{analytic}. Figure \ref{fig:numerical_figures} shows $B$ and $G$ as function of $\theta'$ for various values of $\epsilon$.

Rewriting eq.\eqref{eq:torque3} with eq.\eqref{eq:I}, eq.\eqref{eq:G} and using eq.\eqref{eq:rot} and eq.\eqref{eq:ds}, we obtain by summing over all of the facets $\mathbf{ds}_j$ (assuming $n$ facets):

\begin{equation}\begin{split}
\tau_{x}&=-\sum_{j=1}^{n}\frac{2S_{j}\Phi}{3}\{B(\theta'_{j},\epsilon)\sin\epsilon\sin\theta'_j(x'_j\sin\phi'_j-y'_j\cos\phi'_j)\\
&+G(\theta'_{j},\epsilon)\cos\epsilon\cos\theta'_j(x'_j\sin\phi'_j-y'_j\cos\phi'_j)\}\\
\tau_{y}&=-\sum_{j=1}^{n}\frac{2S_{j}\Phi}{3}G(\theta'_{j},\epsilon)\{\cos\theta'_j(x'_{j}\cos\phi'_j+y'_j\sin\phi'_j)\\
&-z'\sin\theta'_j\}\\
\tau_{z}&=-\sum_{j=1}^{n}\frac{2S_{j}\Phi}{3}\{B(\theta'_{j},\epsilon)\cos\epsilon\sin\theta'_j(x'_j\sin\phi'_j-y'_j\cos\phi'_j)\\
&+G(\theta'_{j},\epsilon)\sin\epsilon\cos\theta'_j(-x'_j\sin\phi'_j+y'_j\cos\phi'_j)\}\\
\label{eq:torque4}.
\end{split}
\end{equation}

We have so far treated the asteroid as having  homogeneous density, so that the center of mass is at the geometric center. Inhomogeneous density
would result in a displacement of the center of mass from the geometric center and can easily be treated
by adding the displacement term, a vector pointing from the geometric CM toward the true CM, to $\mathbf{r'}$ in eq.\eqref{eq:ds}.

%

\subsection{Spin Evolution}
In the inertial system the unit spin vector is:
\begin{equation}\label{eq:spin}
\hat{s}=\begin{pmatrix}
\sin\epsilon\\
0\\
\cos\epsilon
\end{pmatrix}.
\end{equation}

The rate of change of the spin vector's magnitude is:
\begin{equation}\label{eq:spin2}
    \dot{s}=\frac{\boldsymbol{\tau}\cdot\hat{s}}{C}
\end{equation}
where $C$ is the asteroid's moment of inertia.

\begin{figure}[t]
\epsscale{2}
\plottwo{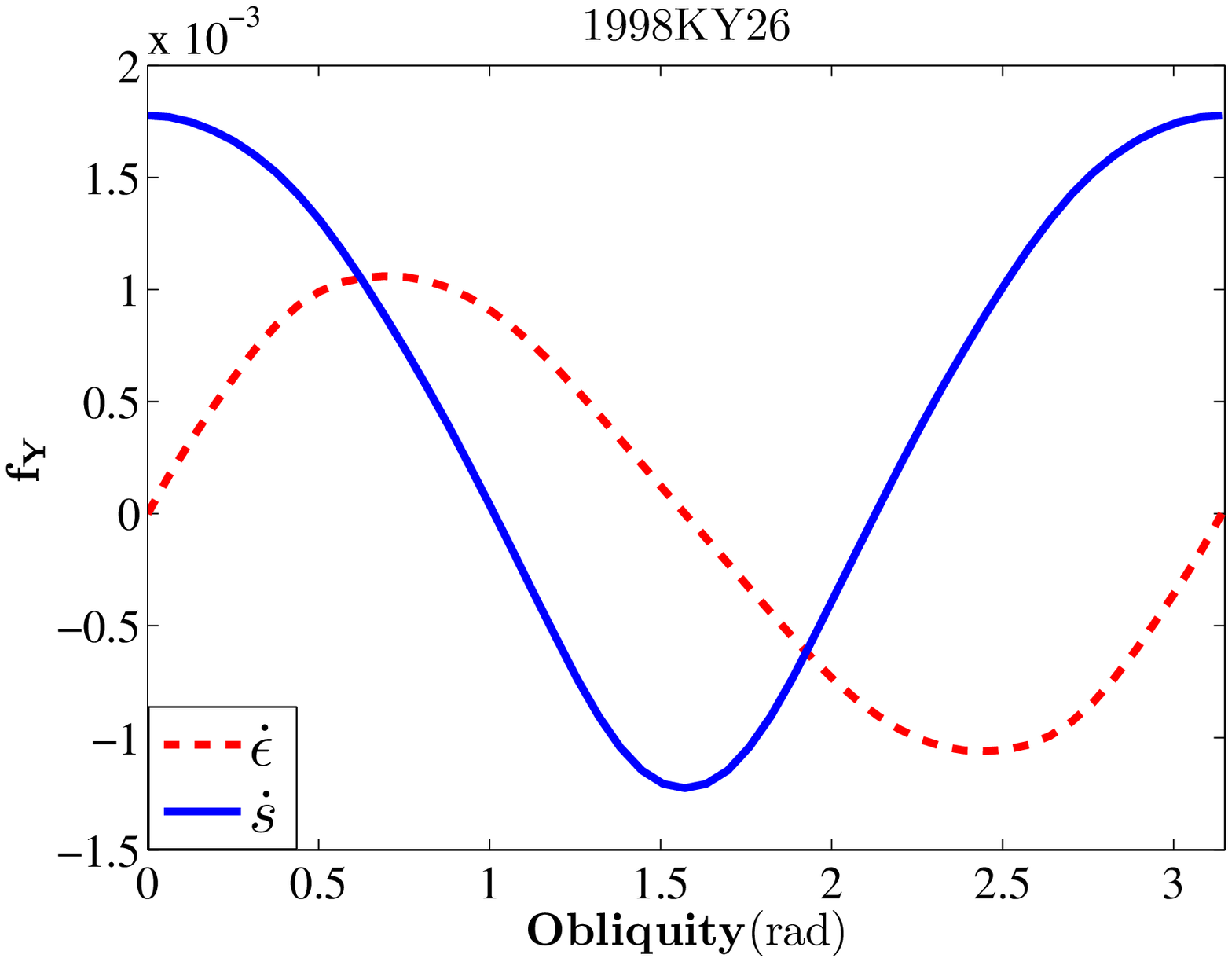}{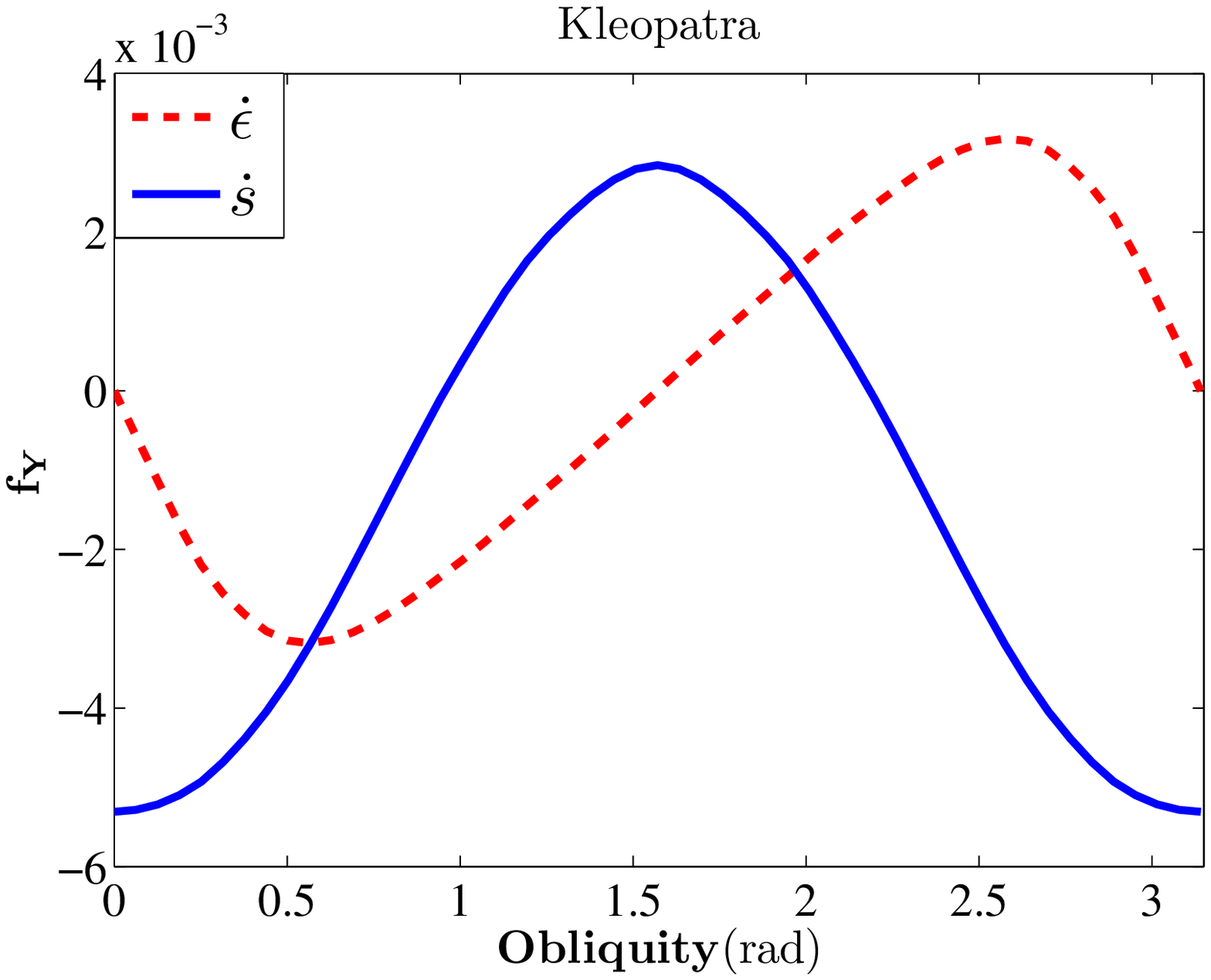}
\caption{The YORP effect calculated for 1998KY26 and Kleopatra. The solid line represents the spin evolution coefficient $f_Y$ and the dashed line represents the obliquity evolution coefficient normalized in the same fashion as $f_Y$. 1998KY26 is a typical Type I asteroid while Kleopatra is a typical Type II asteroid. Zero thermal lag is assumed.}
\label{fig:yorp_calc}
\end{figure}

Substituting eq.\eqref{eq:torque4} into eq.\eqref{eq:spin2} we obtain:
\begin{equation}
  \label{eq:sdot}
   <\dot{s}>=-\sum_{j=1}^{n}\frac{2S_{j}\Phi B(\theta'_{j},\epsilon)\sin\theta'_{j}}{3C}(x'_{j}\sin\phi'_{j}-y'_{j}\cos\phi'_{j}).
 \end{equation}
This result depends only on $B$ since the torque along the $z'$ axis is the only vector that is fixed in all of the coordinate systems. It has the simple physical interpretation as the average insolation times the torque that each facet contributes. Thermal conductivity has no effect on the spin evolution.
An interesting result is that when the obliquity is approximately $\epsilon\approx\ 55^{\circ}$, $B$ is nearly a constant with respect to $\theta'$, and as a result, $\dot{s}$ tends to vanish due to the same argument that was used to show that $\dot{s}$ vanishes due to absorption (refer to \cite{absor}). A clue as to why $B$ is roughly $\theta'$ independent for $\epsilon\approx 55^{\circ}$ might be found in the fact that a related expression: \begin{equation}
\int_{0}^{2\pi}{(\cos\psi\sin\theta')^{2}+(\cos\theta'\sin\epsilon-\cos\epsilon\sin\psi\sin\theta')^{2}}{d}\psi
\end{equation}
is independent of $\theta'$ for $\cos 2\epsilon=-1/3$. This result coincides with the value calculated by \cite{analytic}. Note however, the critical obliquity, $\epsilon\approx 55^\circ$, is by no means a stable point. Even though $\dot{s}=0$, the obliquity will evolve away from the critical obliquity (see \S\ref{sec:obliq}). Our calculations were tested by computing the YORP effect on Kleopatra and 1998KY26, presented in Fig.\ref{fig:yorp_calc}, and comparing it with the results of \cite{CV04} in Fig.3 of their paper for 1998KY26. Our results are about $5\%$ smaller than theirs.

\subsection{Obliquity Evolution}
\label{sec:obliq}
The obliquity change rate can be derived as follows. From eq.\eqref{eq:spin} we have:
\begin{equation}
    \cos\epsilon=\frac{\mathbf{s}\cdot\hat{z}}{s}.
\end{equation}
Differentiating with respect to time we obtain:
\begin{equation}
-\dot{\epsilon}\sin\epsilon=\frac{\dot{\mathbf{s}}\cdot\hat{z}}{s}-\frac{(\mathbf{s}\cdot\hat{z})\dot{s}}{s^{2}}
\end{equation}
and rearranging yields:
\begin{equation}
  \label{eq:obliq2}  \dot{\epsilon}=\frac{\boldsymbol{\tau}[\hat{s}\cos\epsilon-\hat{z}]}{Cs\sin\epsilon}=\frac{\tau_{x}\cos\epsilon-\tau_{z}\sin\epsilon}{Cs}.
\end{equation}
Substituting in eq.\eqref{eq:torque2} we get:
\begin{equation}  \label{eq:obliq}
<\dot{\epsilon}>=-\sum_{j=1}^{n}\frac{2S_{j}\Phi G(\theta'_{j},\epsilon)\cos\theta'_{j}}{3Cs}(x'_{j}\sin\phi'_{j}-y'_{j}\cos\phi'_{j})\Theta
\end{equation}
where $\Theta=1$ for the Rubincam approximation and $\Theta=A$ for the high thermal conductivity regime.
Just like the function $B$ governs the spin's evolution, $G$ governs the obliquity's evolution.
\subsection{Comparison of Qualitative with Precise Results}

\begin{figure}[t]
\plotone{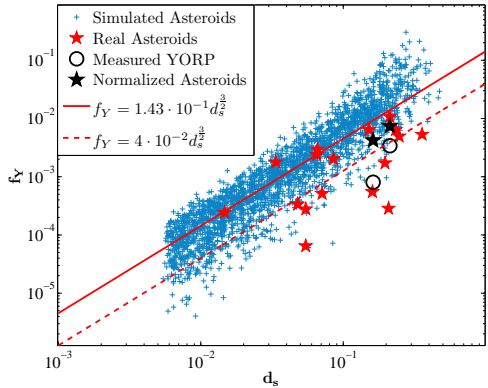}
\caption{Maximal $f_{Y}$ vs. deviation from sphere, the red line denotes the best fit $f_{Y}=0.14\cdot d_{s}^{\frac{3}{2}}$ for the simulated asteroids and the dashed red line denotes the best fit for the real asteroids $f_{Y}=0.04d_{s}^{\frac{3}{2}}$; zero eccentricity is assumed. The filled red stars represent real asteroids calculated with eq.\eqref{eq:sdot}. The filled black stars are the computed $f_Y$ for measured asteroids (normalized to take into account their current obliquity and eccentricity) and the hollow circles are their measured YORP respectively. The asteroids that were calculated are: 216 Kleopatra, 1620 Geographos, 1992 SK, 1998 KY26, 52670 1998 ML14, 2002 CE26, 2053 Bacchus, 4179 Toutatis, 4486 Mithra, 4769 Castalia, 6489 Golveka, 25143 Itokawa, 433 Eros, 1580 Betulia, 54509 YORP (2000 PH5), 1999 KW4 (alpha+beta) and 2100 Rashalom. Zero thermal lag was assumed.}
\label{fig:div}
\end{figure}

In order to check our scaling results of section \S\ref{sec:em}, 2500 random bodies were constructed in the manner described in \S\ref{sec:em}. For each body we computed $f_{Y}$ and $d_s$. Figure \ref{fig:div} shows $f_Y$ for the randomly constructed bodies as a function of spherical deviation, $d_s$, along with the best fit for eq.\eqref{eq:fy}. In addition, $f_Y$ for 18 real asteroids (their shape was taken from ECHO JPL) was computed using our algorithm and plotted together with the 2 measured YORP for comparison. For most of the asteroids we can see a good correlation between our qualitative results and our precise calculations, although our random asteroids tend to have a higher $f_Y$. The ratio between our calculated results and the measured YORP for 1620 Geographos and 2000 PH5 are $2.18$ and $5.22$ respectively. Taking the mean of the real asteroids, we find that $f_{Y}\approx 2.8\cdot10^{-3}$. Reexamining the order of magnitude approach that was taken by GS, we can see that their choice of $f_{Y}~6\times10^{-4}$ is a bit low but is still within acceptable range.

\section{Evolution}
\label{sec:evol}
Since under the mapping $\hat{z}$ to $-\hat{z}$, the obliquity changing component of the torque does not change sign but the $z$ component of the spin vector does, we have the antisymmetry: $G(\theta,\epsilon)=-G(\theta,\pi-\epsilon)$. Similarly, since the spin changing component of the torque does change sign: $B(\theta,\epsilon)=B(\theta,\pi-\epsilon)$.
It can be seen from eq.\eqref{eq:obliq} that every asteroid, regardless of its shape, has at least two equilibrium points with regard to the change in the obliquity since for $\epsilon=0$ or $\pi/2$, $G$ is zero. Whether they are stable or not depends on the derivative of eq.\eqref{eq:obliq} with respect to $\epsilon$, which depends on the asteroid's geometry.
An asteroid will have an identical number of equilibrium points in the range $[0,\pi/2)$ and the range $(\pi/2,\pi]$ due to the symmetry of $G$. The stability of $\dot{\epsilon}$ at $\epsilon=0$ determines the stability of $\dot{\epsilon}$ at all the other equilibrium points since the equilibrium points change stability alternatingly. \cite{VC02} have classified the two most frequent spin and obliquity behaviors of asteroids into Type I and Type II. Type I is categorized by having $\dot{\epsilon}>0$ for $0<\epsilon<\pi/2$ and a spin change rate that is positive for $\epsilon=0$ and negative for $\epsilon=\pi/2$. Type II is the opposite of Type I, categorized by having $\dot{\epsilon}<0$ for $0<\epsilon<\pi/2$ and a spin change rate that is negative for $\epsilon=0$ and positive for $\epsilon=\pi/2$. Both Type I and II are defined to have no nodes for $\dot{\epsilon}$ in the range $(0,\pi/2)$. Our calculations show that $78\%$ of our randomly drawn asteroids are divided into Type I and II with equal likelihood. Our calculations also show that 5 out of the 18 real asteroids are not Type I or II. Of the remaining 13 asteroids, 9 were Type I and 4 were Type II, which is consistent with our random asteroids. One of the curious aspects of both types, is that their spin tends to slow down at their stable point with respect to obliquity. An asteroid which starts with an obliquity in the range of $(0,\pi/2$), will tend to evolve toward $\epsilon=0$ if it is a Type II asteroid where its spin change rate is negative. A Type I asteroid will evolve toward $\epsilon=\pi/2$ where its spin change rate is also negative. Notice that once the spin rate reaches zero, the asteroid does not simply change its obliquity from $\epsilon$ to $\pi-\epsilon$ since that would also involve inverting the $z'$ axis. Rather the obliquity remains fixed while $G$ changes its sign. This causes the equilibrium point to lose its stability and results in the asteroid evolving to its other equilibrium point where the spin will once again slow down.

A viable explanation for this phenomena can be found by inspecting the stability at $\epsilon=0$:\begin{equation}\label{eq:edot_0}
\frac{d\dot{\epsilon}}{d\epsilon}\biggr|_{\epsilon=0}= \sum_{j=1}^{n}\frac{\Phi}{3\pi Cs}\cos^2\theta'_j(S_jx'_j\sin\phi'_j-S_jy'_j\cos\phi'_j).
\end{equation}
The spin change rate with zero obliquity is:\begin{equation}\label{eq:sdot_0}
\dot{s}(\epsilon=0)= -\sum_{j=1}^{n}\frac{2\Phi}{3\pi C}\sin^2\theta'_j(S_jx'_j\sin\phi'_j-S_jy'_j\cos\phi'_j).
\end{equation}

The correlation can be understood as follows. In \S\ref{sec:em} we saw that there is a strong anti-correlation between neighboring facets that causes a reduction in the YORP magnitude. This anti-correlation is represented by the expression in the parentheses in eq.\eqref{eq:edot_0} and eq.\eqref{eq:sdot_0}. To mimic this anti-correlation we take a set, $\{x_n\}$, of N numbers randomly drawn uniformly from -1 to 1 and represent the term in the parentheses as the difference between any two consecutive numbers. For example for the spin change rate we can write:
\begin{equation}
\dot{s}(\epsilon=0)=-\sum_{n=1}^{N-1}(x_{n+1}-x_{n})\sin^2\theta'_n,
\end{equation}
 and for the obliquity stability:
 \begin{equation}
\frac{d\dot{\epsilon}}{d\epsilon}\biggr|_{\epsilon=0}=\sum_{n=1}^{N-1}(x_{n+1}-x_{n})\cos^2\theta'_n
 \end{equation}
where $\theta'$ is a randomly drawn vector of length $N-1$ ranging from $0$ to $\pi$ and sorted in an increasing manner. Since in asteroids we have
\begin{equation}
\sum_{j=1}^{n}S_j\sin\theta'_j(x'_j\sin\phi'_j-y'_j\cos\phi'_j)=0,
\end{equation}
we set $x_{N}-x_{N-1}=\sum_{n=1}^{N-2}(x_{n+1}-x_{n})\sin\theta'_n/sin\theta_{N-1}$.

The sign of $d\dot{\epsilon}/d\epsilon|_{\epsilon=0} \cdot\dot{s}(\epsilon=0)$ for our simple model has a likelihood of $85\%$ being positive, while the likelihood of ${d\dot{\epsilon}}/{d\epsilon}|_{\epsilon=0}\cdot \dot{s}(\epsilon=0)$ for our randomly constructed asteroids being positive is $88\%$ (it is more than $78\%$ because it accounts for cases with more than only 2 stable obliquity points).

The strong tendency of asteroids to spin-down is merely a result of the correlations between neighboring facets.

Asteroids with moderate thermal conductivity might exhibit different obliquity behavior which is beyond the scope of this work \citep{CV04}.

When the spin of the asteroid is sufficiently low, the asteroid might fall into a chaotic tumbling rotation state and it is not clear if the asteroid can recover principal axes rotation \citep{tumbling}.

If the asteroid is a rubble pile, the spin-up might eventually break up the asteroid due to centrifugal force. However, rubble-pile asteroid deformations are not well understood \citep{deform, deform_scheeres, deform_walsh}. Inspection of the shape of binary 1999KW4 \citep{1999kw4} suggests that break up is possible.

\section{BYORP}
The BYORP effect is very similar to the YORP effect so the understanding of YORP continues into BYORP. We will assume that the secondary is synchronously locked with the primary, and that the secondary's spin vector is aligned with the orbital angular momentum. This assumption is reasonable due to the strong tidal interactions between the binary components. The primary is assumed not to be locked with the secondary (i.e., not a double synchronous state), and thus the evolution of the binary orbit does not depend on the orientation of the primary. In the following discussion, the origin of the asteroid system (system 3) will be at the CM of the secondary asteroid with its smallest principle axis, labeled the $x'$ axis, pointing away from the primary. The inertial system's (system 1) origin will now be located at the CM of the primary, the $z$ axis aligned with the orbital angular momentum and the $x$ axis pointing toward the periapsis of the binary's orbit. For the special case where the orbit is circular, we define the $x$ axis to point along the line of the ascending node. The angle $\psi$ is defined relative to the line of nodes where our plane of reference is the orbit of the primary around the sun. In order to be consistent with our previous definitions, the rotation angle $\psi$ is now defined as: $\psi=\lambda+w$, where $\lambda$ is the mean anomaly and $w$ is the argument of periapsis. In addition, the role of the inclination of the binary orbit relative to the primary's orbit around the sun in BYORP takes the role of the obliquity in YORP, $\epsilon\leftrightarrow i$.

In YORP the main interest is to find the evolution of the obliquity and spin. In BYORP we are interested in the evolution of the binary orbit's inclination relative to the orbit of the primary around the sun, the evolution of the binary semi-major axis and the evolution of the binary's eccentricity. Since the secondary is assumed to be locked, changes in the orbital spin vector correspond to change in the semi-major axis, while the change in the inclination corresponds to the change in obliquity in YORP. There is no analog for the eccentricity change in YORP.

Our derivation of BYORP neglects the distance of the secondary's facets from its CM and we assume that all of the facets are located at the CM of the secondary. This is justified since in typical cases the error in the torque due to errors in the facets' locations will  be at least an order of magnitude less than the actual torque.

Just like in YORP, we will inspect the same two extreme of very high and very low thermal conductivity.
In the following section the eccentricity of the orbit of the secondary around the primary will be denoted $\bar{e}$, to distinguish from the eccentricity of the binary orbit around the sun $e$.

Absorbtion of radiation leads to no secular changes in the semi-major axis, eccentricity and inclination of the binary. This can be shown using a similar argument as in \cite{absor}. Fixing the binary's orbital elements, the torque at a given point along the heliocentric orbit cancels out with the torque arising at a point with the true anomaly of the heliocentric orbit changed by $180^\circ$, since the radiation force changes sign. This is true for any point along the orbit, so the net effect of absorption vanishes.

\subsection{Semi-Major Axis Evolution}
\begin{figure}[t]
\plotone{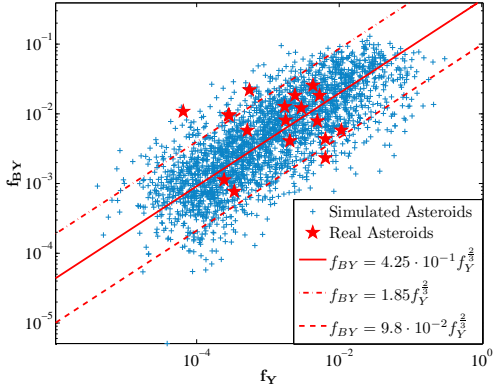}
\caption{The ratio of $f_{BY}$ to $f_{Y}$. The central line corresponds to the best fit of $f_{BY}=0.43f_{Y}^\frac{2}{3}$, while the long dot dashed line represents the upper bound of $f_{BY}=1.85f_{Y}^\frac{2}{3}$ and the short dashed line represents the lower bound of $f_{BY}=0.1f_{Y}^\frac{2}{3}$ of the $2\sigma$ dispersion. Also plotted is the ratio for our real asteroid models, using the same asteroids listed in Fig.\ref{fig:div}. Zero thermal lag is assumed}
\label{fig:fby}
\end{figure}

The change in the semi-major axis can easily be deduced from the change in the energy of a binary orbit. To zero order in the eccentricity the change is:
\begin{equation}\label{eq:adot1}
\frac{1}{a}\dot{a}=\frac{2F_{t}}{na}
\end{equation}
where $a$ is the binary's semi-major axis , $n$ is the mean orbital angular velocity and $F_{t}$ is the transverse orbital force per unit mass arising due to BYORP, positive in the direction of movement. In order to compute $<\dot{a}>$ we define two rotation matrices. The first, $\mathbf{R_{f}}$, describes a rotation around the $z$ axis in the inertial frame by an angle $f+w$, where $f$ is the true anomaly. The second rotation matrix, $\mathbf{R_{\lambda}}$, describes a rotation around the $z$ axis by an angle $\psi$. With our choice of coordinates:
\begin{equation}
F_{t}=-\sum_{j=1}^{n}\frac{2S_{j}}{3M_{s}}\cdot<I>(\mathbf{R_{\lambda}}\cdot\hat{ds'_{j}})\cdot(\mathbf{R_{f}}\cdot\hat{y'})
\end{equation}
here $M_{s}$ is the secondary's mass.
By first averaging over the heliocentric orbit and then averaging over the binary orbit, we obtain:
\begin{equation}\label{eq:adot2}
<F_{t}>=-\sum_{j=1}^{n}\frac{2S_{j}B(\theta'_{j},i)\Phi}{3M_{s}}\sin\theta'_{j}\sin\phi'_{j}.
\end{equation}
This result is intuitive since the expression merely sums the transverse force from each facet. Eq. \eqref{eq:adot2} resembles the YORP spin change rate in eq.\eqref{eq:sdot}. Just like the spin change rate for YORP was unaffected by thermal lag, the semi-major axis change rate is unaffected by thermal lag as well. When $i\approx55^{\circ}$, $<\dot{a}>$ vanishes for the same reasons that $<\dot{s}>$ vanishes. We define $f_{BY}$, similarly to $f_{Y}$, to be the ratio between the BYORP actual effect and the effect it would have on the semi-major change rate if all of the received radiation had been emitted tangentially from the secondary. Therefore, the torque can be rewritten as:\begin{equation}
\boldsymbol{\tau}=\frac{2}{3}\pi aR^2f_{BY}\Phi.
\end{equation}
In the derivation in section \S\ref{sec:em} the angle between the force and the lever that produced the torque scaled like $\mathcal{O}(d_s^\frac{1}{2})$. However, in BYORP there is no correlation between the force and the lever so we would expect:\begin{equation}
f_{BY}\sim d_s\sim f_Y^\frac{2}{3}.
\end{equation}
Fitting the ratio between the coefficients with $f_{BY}=c\cdot f_Y^\frac{2}{3}$ we find that the ratio for our randomly constructed bodies satisfies:
\begin{equation}
f_{BY}=0.43 f_{Y}^\frac{2}{3}.
\end{equation}
The $2\sigma$ dispersion around the best fit number has an upper limit of $f_{BY}=1.85 f_{Y}^\frac{2}{3}$ and a lower limit of $f_{BY}=0.1f_{Y}^\frac{2}{3}$. Figure \ref{fig:fby} shows the ratio for our random asteroids as well as for our real asteroid models. Taking the mean of the real asteroid's $f_{BY}$ yields:
\begin{equation}
f_{BY}\approx 0.01.
\end{equation}

The uncertainties in the asteroid's shape will induce errors in the torque estimate by an amount of:
\begin{equation}
\frac{\Delta \boldsymbol\tau}{\boldsymbol\tau}\approx\frac{2\Delta R}{3(R_s/N)}\approx\frac{2\Delta R}{3 d_s R_s}
\end{equation}
where $\Delta R$ is the physical length scale of the modeling error, $R_s$ is the length scale of the secondary and the factor $2/3$ arises from the relation of YORP to BYORP strength. 

\subsection{Inclination Evolution}

\subsubsection{Zero Order}
The evolution of the inclination can be calculated from \citep[see e.g.][]{Burns76}.
\begin{equation}
\frac{di}{dt}=[a\mu^{-1}(1-\bar{e}^{2})]^{\frac{1}{2}}F_{N}\cos (f+w)/(1+\bar{e}\cos f)
\end{equation}
where $F_{N}$ is the normal component of the BYORP force per unit mass, relative to the binary orbit.
\begin{equation}
F_{N}=-\sum_{j=1}^{n}\frac{2S_{j}}{3M_{s}}\cdot<I>(\mathbf{R_{\lambda}}\cdot\hat{ds'_{j}})\cdot(\mathbf{R_{f}}\cdot\hat{z'})
\end{equation}

Due to the quadrapole moment of the primary, characterized by its $J_2$, the angular momentum of the secondary will be coupled to the angular momentum of the primary \citep{cuk2010}. This interaction causes the orbit of the secondary to rapidly precess, typically at timescales of a few months. This coupling requires us to modify the inclination's evolution rate by a factor $\beta$, the ratio of the secondary's orbital angular momentum relative to the total angular momentum (which may be dominated by the primary's spin). In order to take into account the rapid precession, an averaging of the argument of periapsis is required. No averaging of the line of ascending node is necessary since it does not enter our equations.

The sun will also cause the binary orbit to precess. However, since the timescale for this interaction is of the order of hundreds of years, the angular momentum of the secondary's orbit will stay aligned with the primary's angular momentum. This precession will introduce no additional change in our equations since heliocentric line of nodes does not enter our equations.

The average inclination change rate, up to zero order with respect to the eccentricity is:
\begin{equation}
<\frac{di}{dt}>=-\sum_{j=1}^{n}\frac{2S_{j}G(\theta'_{j},i)\Phi}{3M_{s}na}\Theta\beta\cos\theta'_{j}\sin\phi'_j
\end{equation}
where we added the $\Theta$ term just like in the case for the obliquity evolution.
This result resembles eq.\eqref{eq:obliq}.
\subsubsection{First Order}
In order to simplify our calculations we expand $f$ to first order in the eccentricity, $f\approx \lambda+2e'\sin\lambda$.
To $\mathcal{O}(\bar{e})$ order in eccentricity the contribution is:
\begin{equation}
\begin{split}
<\frac{di}{dt}>&=\sum_{j=1}^{n}\frac{S_{j}\bar{e}\Phi}{3M_{s}na}\cos\theta'_{j}\{3B(\theta'_{j},i)\cos w-\Theta H(\theta'_{j},i)\\
&\times\cos(2\phi'_{j}+w) \}
\end{split}\end{equation}
where the function $H$ is defined to be: \begin{equation}\begin{split}
H(\theta,i)&\equiv\frac{1}{2\pi^{2}}\int_{0}^{2\pi}[\cos2\psi(\cos\psi\sin\theta)^{2}\\
&+(\cos\theta\sin i-\cos i\sin\psi\sin\theta)^{2}]^\frac{1}{2}{d}\psi
\end{split}\end{equation}

Averaging over the precession of the argument of periapsis cancels the first order correction due to eccentricity.

The reasoning presented in \S\ref{sec:evol} leads us to expect a correlation between the change in the inclination and the change in the semi-major axis.
\subsection{Eccentricity Evolution}
Unlike the change in the semi-major axis and inclination which have their analogies in YORP, there is no analogy for the eccentricity evolution.
The eccentricity vector is defined as:
\begin{equation}
\mathbf{\bar{e}}=\frac{\mathbf{v}\times\mathbf{h}}{\mu}-\hat{r}
\end{equation}
where $\mathbf{h}$ is the binary's orbital angular momentum per unit mass, $\mu\equiv  GM_p$ and $\mathbf{v}$ is the orbital velocity of the secondary. Since in an unperturbed orbit this vector is conserved, its time derivative is:
\begin{equation}
\mathbf{\dot{\bar{e}}}=\frac{\mathbf{F}_{BYORP}\times\mathbf{h}}{\mu}+\frac{\mathbf{v}\times\boldsymbol{\tau}_{BYORP}}{\mu}
\label{eq:e'dot}\end{equation}
where $\mathbf{F}_{BYORP}$ is the force that arises from BYORP and $\boldsymbol{\tau}_{BYORP}$ is the torque that is produced by BYORP.
The radius vector $\mathbf{r'}$ in the asteroid system is:
\begin{equation}
\mathbf{r'}=\begin{pmatrix}
\frac{a(1-\bar{e}^2)}{1+\bar{e}\cos f}\\
0\\
0
\end{pmatrix}.
\end{equation}
The time derivative of the eccentricity vector can be written as:
\begin{equation}\begin{split}
\mathbf{\dot{\bar{e}}}&=-\sum_{j=1}^n\frac{2\Phi S_j}{3\mu}<I>\{(\mathbf{R_\lambda}\hat{ds}')\times\mathbf{h}\\
&+\frac{d(\mathbf{R_f}\mathbf{r}')}{dt}\times((\mathbf{R_f}\mathbf{r}')\times(\mathbf{R_\lambda}\hat{ds}'))\}.
\end{split}\end{equation}
Averaging this result over the binary's period, expanding up to $\mathcal{O}(e')$ and taking the projection on the secondary's orbital plane yields:
\begin{equation}
\begin{split}
<\dot{\bar{e}}_{x}>=
&\sum_{j=1}^{n}\frac{S_j\Phi\sin\theta'_{j}}{6M_{s}na}\biggl\{2\Theta G(\theta'_{j},i)[\cos(2\phi'_{j}+w)-3\cos w]\\
&+\bar{e}[2B(\theta'_{j},i)\sin\phi'_{j}-\Theta H(\theta'_{j},i)(\sin(\phi'_{j}+2w)\\
&+3\sin(3\phi'_{j}+2w))]\biggr\}\\
<\dot{\bar{e}}_{y}>=
&-\sum_{j=1}^{n}\frac{S_{j}\Phi \sin\theta'_{j}}{6M_sna}\biggl\{2\Theta G(\theta'_{j},i)[\sin(2\phi'_{j}+w)\\
&-3\sin w)]+\bar{e}[-4B(\theta'_j,i)\cos\phi'_j\\
&+\Theta H(\theta'_{j},i)(3\cos(3\phi'_j+2w)+\cos(\phi'_j+2w))]\biggr\}\\
\end{split}
\end{equation}

\subsubsection{Eccentricity's Magnitude Evolution}
The change in the magnitude of the eccentricity is the projection of $\mathbf{\dot{\bar{e}}}$ on the $x$ axis. To zero order in eccentricity this change is:
\begin{equation}
<\dot{\bar{e}}>=\sum_{j=1}^{n}\frac{S_jG(\theta'_{j},i)\Phi\sin\theta'_{j}}{3M_{s}na}\Theta[\cos(2\phi'_{j}+w)-3\cos w].\end{equation}
This contribution is zero for $i=0,\pi/2$.

Up to $\mathcal{O}(\bar{e})$ the change is:
\begin{equation}
\begin{split}
<\dot{\bar{e}}>&=\sum_{j=1}^{n}\frac{S_j\Phi\sin\theta'_{j}}{6M_{s}na}\bar{e}[2B(\theta'_{j},i)\sin\phi'_{j}-\Theta H(\theta'_{j},i)\\
&\times(\sin(\phi'_{j}+2w)+3\sin(3\phi'_{j}+2w))].
\end{split}\end{equation}
Averaging over the precession of the argument of periapsis retains only the term:
\begin{equation}
\label{eq:edot1}
<\dot{\bar{e}}>=\sum_{j=1}^{n}\frac{S_j\Phi}{3M_{s}na}\bar{e}B(\theta'_{j},i)\sin\theta'_{j}\sin\phi'_{j}.
\end{equation}
Comparing eq.\eqref{eq:adot1} and eq.\eqref{eq:adot2} to eq.\eqref{eq:edot1} we find, similarly to GS, that  the eccentricity evolves like:
\begin{equation}
\bar{e}\propto a^{-\frac{1}{4}}.
\end{equation}
We recover the result of \cite{byorp_1999kw4} that circular orbits remain circular. An interesting result is that both the semi-major axis evolution and eccentricity evolution are independent of the asteroid's albedo and thermal conductivity, in the two extreme regimes of high and low thermal conductivities.

 \subsection{Results}
\begin{figure}[t]
\plotone{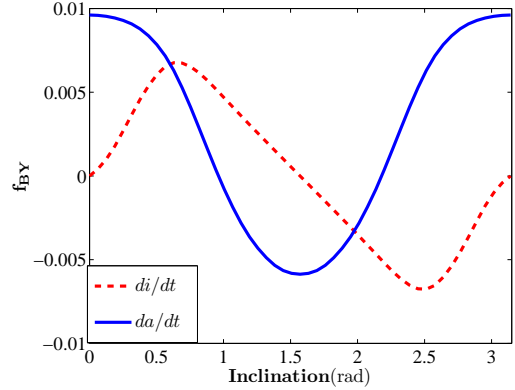}
\caption{The BYORP effect calculated for 1999KW4. The solid line is the semi-major axis changing coefficient $f_{BY}$ and the dashed line represents the inclination changing torque coefficient normalized in the same fashion as $f_{BY}$. 1999KW4 is a typical Type II asteroid, zero thermal lag is assumed.}
\label{fig:1999kw4}
\end{figure}

Our calculations were tested on the binary asteroid 1999KW4 which is the best modeled binary. Figure \ref{fig:1999kw4} shows our calculated change in the semi-major axis and in the inclination of the binary. Our calculations show that the secondary is currently drifting away from the primary at a rate of about $\dot{a}=7\; \text{cm}/\text{year}$, which is in an agreement to within $5\%$ with the result of \cite{byorp_1999kw4}.

Due to the similarity between YORP and BYORP, the statistics found in \S\ref{sec:evol} are similar for BYORP. We preformed calculations on the 2500 randomly constructed asteroids. The stability of $di(i=0)/dt$ with respect to a change in the inclination determines the stability of the rest of the equilibrium points. The correlation that was found between the sign of $\dot{s}(\epsilon=0)$ and the stability at $\epsilon=0$ also holds for the sign of $\dot{a}(i=0)$ and the stability at $i=0$. For randomly shaped asteroids, there is an equal likelihood for $i=0$ to be a stable point or an unstable point and $73\%$ were either Type I or Type II (with equal likelihood). Our calculations also show that 2 out of the 18 real asteroids are not Type I or II. Of the remaining 16 asteroids, 6 were Type I and 10 were Type II, which is consistent with our random asteroids.

Unlike YORP, the timescale for reaching the stable point can be longer than the lifetime of the system, so not every binary will reach it. In order for the system to reach its stable point, the difference between the starting inclination and the inclination at the stable point needs to be $\Delta i\approx \beta^{-1}$.

\section{Conclusions}
In this paper we have developed an intuitive and simple analytical model that predicts the behavior of the YORP and BYORP effects. Our model was computed on randomly constructed asteroids and on shape models of real asteroids. The calculations of our simple model show that the YORP effect is dominated by the inner correlations between facets. These correlations explain the magnitude of the effect and its preference to spin-down the asteroid. We have also shown that the dimensionless parameter of $d_s$ governs the strength of the effect (but not of the sign).

Kilometer sized or smaller near earth asteroids undergo a relative fast spin-up or spin-down due to YORP. The YORP effect tends, in most cases, to change the initial obliquity of the asteroid to either $\epsilon=0$ or $\epsilon=\pi/2$. For both extreme regimes of thermal conductivity asteroids will most likely spin-down. The rapid spin-up of asteroids, will eventually cause the asteroid to break up and possibly form a binary. The binary will evolve under BYORP and tidal forces; the former will tend to orient the binary into a $i=0$ or $i=\pi/2$ orbit, if the system is near enough relative to the stable point. After the binary orbit reaches its stable inclination, the binary will most likely migrate inwards until it is stopped by tidal forces (if it survives the migration). A simple relation between the magnitude of the YORP effect and BYORP effect is calculated, thus knowledge of the former can help us estimate the strength of the latter.

\begin{acknowledgments}
We would like to thank Lance Benner for providing us the shape models for the real asteroids. We would also like to thank the referee Matija \'{C}uk for alerting our attention about the importance of the $J_2$ interactions.
\newline
\end{acknowledgments}

\bibliography{stei031010}

\end{document}